# Augmented two-stage estimation for treatment crossover in oncology trials: Leveraging external data for improved precision

Harlan Campbell[1,2], Nicholas Latimer[3], Jeroen Jansen[1,4], Shannon Cope[1]

1. Precision AQ, Evidence Synthesis and Decision Modeling, Vancouver, BC, Canada
2. Department of Statistics, University of British Columbia, Vancouver, BC, Canada
3. School of Health and Related Research, University of Sheffield, Sheffield, UK
4. School of Pharmacy, University of California – San Francisco, CA, United States

**ABSTRACT**

Randomized controlled trials (RCTs) in oncology often allow control group participants to crossover to experimental treatments, a practice that, while often ethically necessary, complicates the accurate estimation of long-term treatment effects. When crossover rates are high or sample sizes are limited, commonly used methods for crossover adjustment (such as the rank-preserving structural failure time model, inverse probability of censoring weights, and two-stage estimation (TSE)) may produce imprecise estimates. Real-world data (RWD) can be used to develop an external control arm for the RCT, although this approach ignores evidence from trial subjects who did not crossover and ignores evidence from the data obtained prior to crossover for those subjects who did. This paper introduces "augmented two-stage estimation" (ATSE), a method that combines data from non-switching participants in a RCT with an external dataset, forming a "hybrid non-switching arm". With a simulation study, we evaluate the ATSE method's performance compared to TSE crossover adjustment and an external control arm approach. Results indicate that performance is dependent on scenario characteristics, but when unconfounded external data are available, ATSE may result in less bias and improved precision compared to TSE and external control arm approaches. When external data are affected by unmeasured confounding, ATSE becomes prone to bias, but to a lesser extent compared to an external control arm approach.

## 1 Introduction

Randomized controlled trials (RCTs) in clinical research often include the option for participants randomized to the control treatment (i.e., placebo or standard of care) to crossover to the experimental treatment after a predefined time. This practice appears increasingly common in oncology and, from an ethical perspective, helps ensure that trial participants are not denied access to potentially beneficial new treatments. Allowing for crossover is also thought to improve recruitment/enrolment [1, 2], (but see Chen and Prasad who conclude otherwise[3]). Unfortunately, treatment crossover (or "switching") can complicate the estimation of overall survival benefits, as decision-makers are often interested in the hypothetical treatment effect of the experimental treatment versus the control treatment in jurisdictions where the experimental therapy is not yet available in the treatment pathway as a later line of therapy. [4]

In oncology trials, where progression-free survival (PFS) is frequently the primary endpoint, treatment crossover in the control group upon disease progression can obscure the estimation of overall survival (OS) -- a key measure for reimbursement and health technology assessments (HTA). To address these challenges, statistical adjustment methods have been proposed, such as rank-preserving structural failure time models (RPSFT)[5], inverse probability of censoring (IPCW)[6], and two-stage estimation (TSE)[7]. These approaches improve upon naïve adjustment methods that simply censor subjects who switch treatments, given that switchers often have a different prognosis than non-switchers. However, these crossover adjustment methods often result in highly uncertain estimates, especially in studies with a high rate of crossover and/or a small sample size.

The National Institute for Health and Care Excellence (NICE) Decision Support Unit recently published a Technical Support Document (TSD 24, April 2024) that summarizes recommendations for adjusting survival time estimates in the presence of treatment switching in clinical trials. TSD 24 highlights the potential of using external data for crossover adjustment following similar calls in earlier work[8 9 10]. However, until now, external data has only been used for crossover adjustment in select examples, where external evidence was used to: 1) construct an external control arm (ECA) that was not impacted by switching and replace the RCT control arm considering alignment with the target trial population[11]; 2) validate and select between different crossover adjustment methods (IPCW, TSE, and RPSFTM)[12]; 3) estimate what post-progression survival (PPS) would have been in the RCTs had treatment switching not occurred.[13] This final example was the first to integrate external data into a TSE crossover adjustment model.[13] Although the Evidence Review Group report[14] raised concerns with regards to the use of external data in this case,[*] Latimer and Abrams[15] noted that "the deliberations of the [NICE] Appraisal Committee regarding TA171 demonstrated openness to the use of external data in the presence of treatment switching."

[*]First, it was noted that there were no analyses done to understand the likelihood of unmeasured confounders (i.e., to determine if it was possible to fully adjust he external trial). Second, there were concerns that the external data were dated and possibly no longer relevant to the trial due to a long-term trend of improving survival. Finally, the adjusted treatment estimates seemed rather implausible (i.e., the results suggested that receiving treatment after disease progression had the same (if not higher) benefit on overall survival (OS) then when received before disease progression).

In this paper, we propose an alternative way of using external data for crossover adjustment based on the concept of a "hybrid control arm". Methods for hybrid-control arm studies are not new, but are only recently being considered in oncology research[16] (e.g., using a control arm formed from historical clinical trials in metastatic colorectal cancer[17] or from EHR-derived data, such as the Flatiron Health database for advanced non-squamous non-small cell lung cancer (aNSCLC)[18]). Our proposed augmented two-stage estimation (ATSE) method leverages a "hybrid non-switching arm" for comparison with the RCT switching arm, enabling more efficient estimation of survival beyond the switching time-point. We detail the ATSE method in Section 2, present a simulation study in Section 3, and conclude with a discussion in Section 4.

# 2 Methods

We begin by defining basic notation and summarizing the standard two-step estimation (TSE) procedure. We then describe the augmented two-stage estimation (ATSE), which uses a hybrid non-switching arm to estimate the effect of treatment switching on survival.

## 2.1 Notation

Suppose subjects in the RCT are randomly allocated to either the "control arm" or the "treatment arm", with the primary endpoint of interest being right censored OS. Subjects who are randomized to the control are allowed to switch from the control to the experimental treatment following disease progression and we assume that any switches are likely to happen immediately or shortly following progression. Suppose also that patient-level data from an external cohort is available for subjects treated with the control who did not switch onto the experimental treatment. Finally, suppose a decision-maker is interested in evaluating the hypothetical OS benefit of the treatment versus the control in a jurisdiction where the experimental treatment is not available in subsequent lines of therapy.

The $i$-th subject contributes $S_i$, $A_i = (A_{1i}, A_{2i})$, $W_i$, $T_i = (PFS_i, PPS_i)$, $C_i = (PFC_i, PPC_i)$, and $\boldsymbol{X_i}$. Let $S_i$ denote trial participation status, with $S_i = 1$ indicating that the $i$-th subject is in the RCT and $S_i = 0$ indicating that the $i$-th subject is in the external data. Let $A_i = (A_{i1}, A_{i2})$ correspond to the $i$-th subject's sequence of binary treatment assignments (with 0 corresponding to control and 1 corresponding to treatment). Let $W_i$ be a binary indicator corresponding to treatment switch (with $W_i = 1$ indicating switch and $W_i = 0$ indicating no switch). For example, $A_i = (0,1)$ implies $W_i = 1$ and indicates that the $i$-th subject was initially assigned/randomized to the control and then, following progression, switched to the experimental treatment. We assume that $S_i = 0$ implies that $A_i = (A_{i1}, A_{i2}) = (0, 0)$ since all subjects in the external data are "assigned" to the control and do not switch (switching is only permitted for those in the RCT initially randomized to the control arm). Let $T_i = (PFS_i, PPS_i)$ correspond to the $i$-th subject's PFS and PPS times, with $C_i = (PFC_i, PPC_i)$ indicating if these are observed or censored. Finally, $\boldsymbol{X}_i$ corresponds to a vector of covariates measured at progression (or assumed to be fixed from baseline). Say $\boldsymbol{X_i}$ partitions as variables $\boldsymbol{X_{1i}}$ which are common causes of post-progression survival ($PPS_i$) and switching ($W_i$), and as variables $\boldsymbol{X_{2i}}$ which are common causes of post-progression survival ($PPS_i$) trial participation status (i.e., participation in RCT versus the external study) ($S_i$).

**2.2 Two-stage estimation (TSE)**

As its name suggests, the TSE approach for crossover adjustment involves two main steps. First, one estimates the effect of treatment switching on PPS. Then, in the second step, this effect estimate is used to estimate counterfactual survival times that would have been observed if switching had not occurred.

**Step 1.** In the first step, a standard parametric accelerated failure time (AFT) model (such as a Weibull model) is fit to all subjects randomized to the control arm of the RCT, relating PPS to switching ($W$), and all possible confounders ($\boldsymbol{X_1}$) (i.e., all variables that are common causes for post-progression survival and switching). For example, a Weibull model could be specified such that:

$$PPS_i = \exp(\alpha + \mu W_i + \boldsymbol{X_{1i}^t}\gamma + \sigma\epsilon_i), \quad (1)$$

for the $i$-th subject (for all subjects for which $S_i = 1$ and $A_{i1} = 0$), where $\alpha$ is the intercept parameter, $\mu$ is a parameter corresponding to the effect of switching, $\gamma$ is a parameter-vector corresponding to the effect of the possible confounders, $\epsilon_i$ is an error term that has the extreme value distribution, and $\sigma$ is the scale parameter. In an AFT model the covariates act multiplicatively on time. For example, suppose the acceleration factor is $\exp(\mu) = 4$, then the effect of switching would be to quadruple the PPS.

**Step 2.** After obtaining parameter estimates from the AFT model, estimated counterfactual PPS times, $\widehat{U_i}$, are obtained as follows:

$$\widehat{U_i} = W_i \frac{PPS_i}{\exp(\hat{\mu})} + (1 - W_i)PPS_i, \quad (2)$$

for the $i$-th subject (for all subjects for which $S_i = 1$ and $A_{i1} = 0$); where $\exp(\hat{\mu})$ is the estimated acceleration factor associated with switching obtained from the fitted AFT model in Step 1. Note that no adjustment is made for subjects who do not switch (i.e., if $W_i = 0$, then $\widehat{U_i} = PPS_i$).

Suppose for example that the effect of switching is to double the PPS (i.e., $\exp(\hat{\mu}) = 2$). Then for subjects that switched, their counterfactual PPS times are obtained by halving their observed PPS times. Adjusted overall survival (OS) times are then obtained by adding the counterfactual PPS times to the observed PFS times: $\widehat{OS_i} = PFS_i + \widehat{U}_i$. If censoring is present, an additional step called "re-censoring" can be conducted; see Zhang and Chen (2016) [19] for details and see Latimer et al. (2019)[20] for a discussion about the appropriateness of this step.

After estimating the untreated survival times for patients who switched treatments (and conducting any necessary re-censoring), a new "adjusted RCT" dataset is created. This dataset combines observed OS times for patients who did not switch treatments with adjusted OS times for those who did. This adjusted RCT dataset can then be used to estimate the relative treatment effect with respect to OS. For example, a Cox proportional hazards model could be fit to estimate a crossover-adjusted hazard ratio (HR). Alternatively, the relative treatment effect could be estimated non-parametrically as a crossover-adjusted difference in restricted mean survival time (dRMST). Valid confidence intervals for these estimates can be obtained by bootstrapping the entire adjustment and estimation process.

Note that to obtain unbiased results, the TSE method relies on the assumption of "no unmeasured confounders". This means that the Step 1 AFT model must include all variables that predict both PPS and treatment switching. In other words, $\boldsymbol{X_1}$ must include all prognostic variables that could influence a participant's decision of whether to switch (or their doctor's decision to recommend that they switch). It may be difficult to determine which variables influence a decision to switch, and ultimately this assumption of "no unmeasured confounders" cannot be tested empirically. However, clinicians may provide valuable information about treatment switching decisions and directed acyclic graphs (DAGs), such as the Figure 1 diagram, may be a useful tool to help inform covariate selection. [21]

The DAG in Figure 1 can be understood to illustrate the causal relationships between each of the variables. For instance, the arrow from $\boldsymbol{X_1}$ to $\boldsymbol{W}$ implies that the variables included in $\boldsymbol{X_1}$ directly influence a decision to switch and the arrow from $\boldsymbol{X_1}$ to ***PPS*** implies that these variables also directly impact the post-progression survival. For example, "disease severity" as measured shortly before progression might be included in $\boldsymbol{X_1}$ and would influence the decision to switch (e.g., those with high severity are recommended by their doctor's to not switch to the experimental treatment) and the duration of PPS.

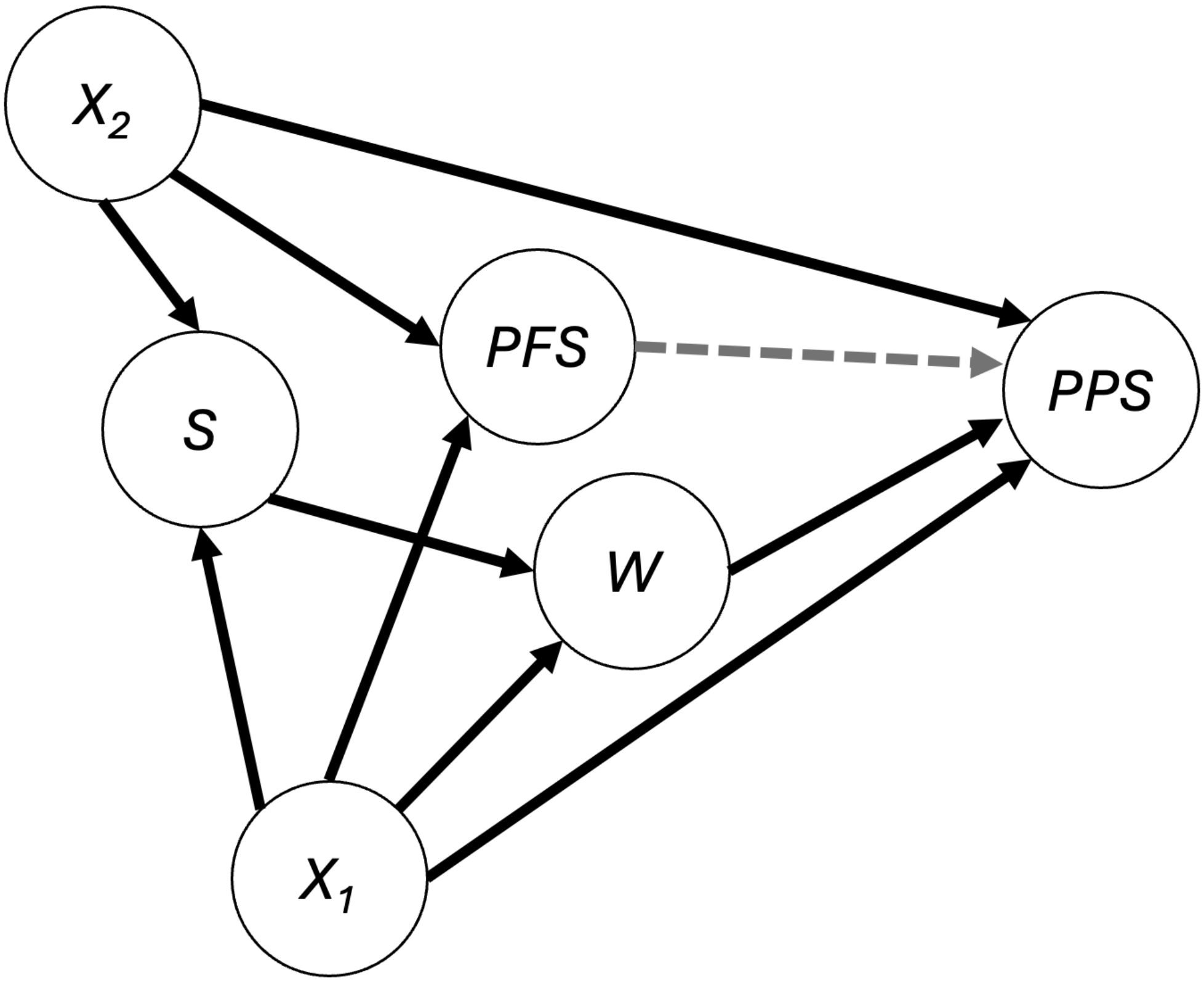

**Figure 1 –** For estimating the effect of switching (*W*) on post-progression survival (*PPS*), there are two minimal sufficient adjustment sets: $\{S, X_1\}$ and $\{X_1, X_2\}$. The AFT model fit in Step 1 of the TSE is restricted to only those individuals with $S = 1$ (i.e., only subjects in the RCT) and must therefore include $X_1$ (i.e., all variables that predict both PPS and treatment switching). The AFT model in Step 2 of the ATSE is fit to individuals from both the RCT ($S = 1$) and external data ($S = 0$) and must therefore include $X_1$ and $X_2$ (i.e., all variables that predict both PPS and treatment switching as well as all variables that predict both PPS and trial participation status).

### 2.3 Augmented two-step estimation (ATSE)

For TSE, limited sample size in the non-switching subjects in the control arm of an RCT increases uncertainty in the estimation of the acceleration factor (i.e. $\exp(\mu)$), leading to an imprecise estimate of the counterfactual survival times and ultimately the crossover-adjusted relative treatment effect. Leveraging external data for subjects assigned to the control treatment who do not switch (e.g., from historical RCTs or RWE) could reduce the uncertainty in the estimation of the acceleration factor, resulting in a more precise crossover adjustment.

Various methods have been developed to construct so-called "hybrid control arms" whereby a trial's small control arm is combined with individuals from an external data source. These methods are mostly

Bayesian approaches and include (modified) power prior models (Ibrahim and Chen, 2000[22]), commensurate prior models (Hobbs et al., 2012[23]), and robust meta-analytic predictive prior (RMAPP) models (Schmidli et al., 2014[24]). Most recently, the Bayesian latent exchangeability prior (LEAP) model (Alt et al., 2024[25]) appears particularly promising (Campbell and Gustafson, 2024[26]). However, since the TSE method is frequentist, for the ASTE we adopt the two-step dynamic borrowing approach recently proposed by Tan et al. (2022)[27], which reflects a frequentist analog to the modified power prior method.

The augmented TSE method consists of the following four steps.

**Step 1.** In the first step, a standard parametric AFT model (such as a Weibull model) is fit comparing the PPS between those randomized to the control arm who did not switch and those in the external control arm, adjusting for covariates $\boldsymbol{X} = \{\boldsymbol{X_1}, \boldsymbol{X_2}\}$. To be clear one must adjust for all common causes of post-progression survival and switching (i.e., adjust for $\boldsymbol{X_1}$), as well as all common causes of post-progression survival and trial participation status (i.e., adjust for $\boldsymbol{X_2}$); see Figure 1.

For example, a Weibull model could be specified such that:

$$PPS_i = \exp(\beta + \rho S_i + \boldsymbol{X}_i^t \eta + \tau \epsilon_i), \tag{3}$$

for the $i$-th subject (for all subjects for which $W_i = 0$ and $A_{i1} = 0$), where $\beta$ is the intercept parameter, $\epsilon_i$ is an error term that has the extreme value distribution, and $\tau$ is the scale parameter. The value of the $\rho$ parameter corresponds to the degree of dissimilarity between the two cohorts. Specifically, $\exp(\rho)$ is the acceleration factor associated with participation in the RCT versus the external cohort (after adjusting for all measured confounders). Suppose for example that $\exp(\rho) = 2$. This would indicate that individuals in the RCT have PPS times that are twice as long on average as individuals with similar values of $\boldsymbol{X}$ who are in the external data cohort. A large value of $|\rho|$ would therefore indicate a large degree of dissimilarity between cohorts that cannot be attributed to measured confounders and would suggest the possibility of substantial unmeasured confounding.

Dynamic borrowing then considers the degree of dissimilarity in determining how much information to borrow from the external cohort. If the magnitude of $\rho$ is small, this suggests that the external data is relatively compatible with the RCT data. On the other hand, if the magnitude of $\rho$ is large, this suggests that the external data is less compatible with the RCT data and should therefore be down-weighted in the analysis. To calculate the amount of cohort-level downweighing as a function of $\rho$, Tan et al. (2022) [27] consider an exponential function which assigns each individual in the external data a weight equal to

$$\widehat{w}_i = \exp(-c|\hat{\rho}|), \tag{4}$$

for the $i$-th subject (for all subjects for which $S_i = 0$); where $c > 0$ is a pre-specified constant "decay factor", akin to the power parameter in the Bayesian modified power prior approach. A larger value of $c$ will result in borrowing less information from the external data (i.e., in a faster decay to 0 as the difference between the RCT and external cohorts increases). All individuals in the RCT (all subjects for which $S_i = 1$) are given a weight equal to 1.

**Step 2.** The second step of the ATSE method is to fit a second AFT model to the weighted control subjects (i.e., all subjects with $A_{i1} = 0$) with the weights defined from the previous step. This second AFT model relates PPS to switching ($W$), and includes covariates $\boldsymbol{X} = \{X_1, X_2\}$, the same covariates which were included in the first step AFT model:

$$PPS_i = \exp(\alpha + \mu W_i + \boldsymbol{X}_i^t \gamma + \sigma \epsilon_i), \qquad (5)$$

for the $i$-th subject (for all subjects for which $A_{i1} = 0$), where $\alpha$ is the intercept parameter, $\epsilon_i$ is an error term that has the extreme value distribution, and $\sigma$ is the scale parameter. This is similar to the first step of the TSE approach, except that the dataset here is expanded to include the down-weighted external data, rather than only control patients of the RCT.

**Step 3.** The third step of the ATSE approach is deriving the estimated counterfactual PPS times, $\widehat{U_i}$, and OS times, $\widehat{OS_i}$ (for all subjects with $S_i = 1$ and $A_{i1} = 0$). This is done as in TSE (see equation (2)) based on $\exp(\hat{\mu})$, the estimated acceleration factor associated with switching obtained from fitting model (5) in the second step. If censoring is present, "re-censoring" can be applied; see Latimer et al. (2019)[20].

**Step 4.** The fourth and final step of the ATSE approach is to estimate the relative treatment effect of interest based on a new "adjusted RCT" dataset which combines observed OS times for RCT patients who did not switch treatments with adjusted OS times for those RCT patients who did switch. For instance, a Cox proportional hazards model, or an AFT model, relating OS to treatment at randomization ($A_1$) could be fit with the adjusted RCT dataset to estimate the relative treatment effect in terms of a hazard ratio (HR) or an acceleration factor (AF). As with TSE, valid confidence intervals can be obtained by bootstrapping the entire adjustment and estimation process.

It is important to reiterate that beyond the "no unmeasured confounders" assumption required in TSE (i.e., one must adjust for all variables that predict both PPS and treatment switching), the ATSE method requires an additional "no unmeasured confounders" assumption with respect to the external data. The DAG in Figure 1 illustrates that there are two minimal sufficient adjustment sets for estimating the effect of switching ($W$) on post-progression survival ($PPS$): $\{S, X_1\}$ and $\{X_1, X_2\}$. The AFT model fit in Step 1 of the TSE is restricted to only those individuals with $S = 1$ (i.e., only subjects in the RCT) and must therefore include $X_1$ (i.e., all variables that are common causes of $PPS$ and $W$). The AFT model in Step 2 of the ATSE is fit to individuals from both the RCT ($S = 1$) and external data ($S = 0$) and must therefore include $X_1$ and $X_2$. Down-weighting the external data according to the suspected amount of unmeasured confounding as determined by the AFT model in Step 1 of the ATSE will help reduce the impact of confounding bias but cannot entirely eliminate it. With regards to covariate selection, a conservative approach would be to adjust for all covariates that are considered important prognostic factors.

We have included ***PFS*** in the Figure 1 DAG with a dashed line towards ***PPS*** to indicate a potential direct causal link. Regardless of whether there is a direct causal link, ***PFS*** might serve as a useful "proxy confounder" in the sense that it might be highly correlated with $\boldsymbol{X_1}$ and/or $\boldsymbol{X_2}$. For instance, if "disease severity" measured shortly before progression is deemed a confounder within $\boldsymbol{X_1}$ or $\boldsymbol{X_2}$, but is unmeasured in the data, one might consider ***PFS*** as a proxy for this unmeasured confounder (since those with high "disease severity" will likely have shorter progression times and vice-versa). VanderWeele (2019) explains that "in most cases adjustment for such variable will reduce the bias due to confounding". [28] In order to adjust for PFS in the ATSE one would simply expand the covariate set $\boldsymbol{X}$ to include *PFS* in Step 1 and Step 2: $\boldsymbol{X} = \{\boldsymbol{X_1}, \boldsymbol{X_2}, \boldsymbol{PFS}\}$. In the Section 3 simulation study, we will consider the merits of this strategy.

# 3 Simulation Study

## 3.1 Objectives

To evaluate the performance of the ATSE method, we conducted a simulation study designed to mirror typical settings in clinical research with treatment switching and incorporation of external data. This simulation study aimed to assess the accuracy and robustness of the ATSE approach under various conditions, including variations in switching rates, and different decay factors for the ATSE hybrid arm. Here, we provide a summary of the data generating mechanism, the estimand of interest, and the methods we compared according to various performance measures. The simulation study was conducted using R. The code used to simulate the data is provided in the Appendix.

## 3.2 Data generation and mechanism

We followed data generation procedure used outlined in Latimer et al. (2020)[29] for their eight "simple scenarios" with a few modifications. Specifically, we simulated RCT datasets with a sample size of $N_{RCT} = 500$ and 2:1 randomization in favour of the experimental group, and with treatment switching permitted from the control group to the experimental treatment following progression.

Overall survival times were simulated based on a 2-component mixture Weibull baseline survival function and were dependent on three binary variables: treatment, prognosis ($badprog$), and an unmeasured prognostic factor ($U$). (Note the simulation study by Latimer et al. (2020)[29] does not include a dependency on $U$, but is otherwise the same). The corresponding hazard function is:

$$h_i(t) = h_0(t)\exp(\delta_1 trt_i + 0.3badprog_i - 0.3U_i), \quad (6)$$

for the $i$-th subject; where $h_0(t)$ represents the baseline hazard function and $\delta_1$ represents the log hazard ratio (log-HR) associated with treatment. The bad prognosis and unmeasured confounder variables were simulated as independent binary variables such that each individual had a 50% probability of a bad prognosis and a 50% probability of $U_i$=1. We assumed no treatment effect heterogeneity (i.e., no effect modifiers/interaction terms). Also, while in reality certain covariates, such as prognosis, might change (worsen) over the course of the study (e.g., a subject's disease severity may be notably different at baseline than at progression), we assumed that all covariates remain fixed over the course of the study.

PFS times were simulated as a function of OS times, so that on average an individual's PFS was one third of their OS. More specifically, PFS times were equal to OS times multiplied by a random draw from a Beta(5,10) distribution. A short delay was assumed between an individual's true progression time and their observed progression time by setting the observed progression time as equal to their first "visit time" following the progression event, with "visits" simulated every 21 days from randomisation to death.

We simulated external control datasets in the same way that the RCT datasets were simulated but with $trt_i$=0 for all subjects. Also, whereas individuals in the RCT had 50% probability of bad prognosis, individuals in the external control dataset had a 75% probability of bad prognosis.

Sixteen scenarios were simulated varying (1) the magnitude of the treatment effect, (2) the degree of crossover, (3) the censoring, and (4) the sample size of the external control arm; see Table 1. These scenarios were based on Latimer et al. (2020)[29]'s eight simple scenarios. Specifically, for scenarios with "low treatment effect", we set $\delta_1$ = -0.2, whereas for scenarios "high treatment effect", we set $\delta_1$ = -0.5. For scenarios with "moderate switching", RCT individuals with poor prognosis had an 80% probability of switching, whereas those with a good prognosis group had a 30% probability of switching. For scenarios with "high switching", the probability of switching for those in the control arm was set to 90% in the poor prognosis group and to 60% in the good prognosis group. To be clear, in this simulation study, all individuals in the external control and all those in the experimental treatment arm of the RCT do not switch treatments. In the scenarios with "no censoring", individuals were only censored if still alive at 5000 days following randomization (the end of the study), whereas scenarios with "moderate censoring" censored all subjects at 546 days following randomization (the end of the study). Finally, the sample size of the external control arm was either $N_{EC}$=200 (resembling historical trial data) or $N_{EC}=1000$ (resembling registry data).

To investigate the impact of unmeasured confounding, for each of the sixteen scenarios, 2000 datasets were simulated with three different conditions. In Condition A there was no unmeasured confounding, meaning the unmeasured prognostic factor $U$ did not impact the probability of switching and had the same distribution in the RCT and external control (i.e., all subjects in the RCT and external control, Pr $(U_i = 1) = 0.5$). In Condition B confounding bias in the external control was considered by setting the prognostic factor $U$ in the RCT to Pr $(U_i = 1) = 0.50$ and in the external control to Pr $(U_i = 1) = 0.75$. Finally, in Condition C, confounding bias in the crossover-adjusted control arm of the RCT was considered, where the unmeasured prognostic factor $U$ impacted crossover, reducing the probability of switching by 20% for all individuals in the RCT for whom $U_i = 1$. In this case, the external data remains unbiased with Pr $(U_i = 1) = 0.50$ for all subjects in both the RCT and the external data. Note that the same random number seed was used for all three conditions within each scenario so that the RCT data simulated in Condition A and Condition B will be identical and the external data simulated in Condition A and Condition C will be identical.

### 3.3 Estimands of interest

As in Latimer et al. (2020)[29], the estimand of interest for the simulation study was the restricted mean survival time (RMST) in the control group at study end-date (5000 days for Scenarios 1-4 and 9-12; 546 days for Scenarios 5-8 and 13-16). The true RMST value in the control group was 472.75 days in Scenarios 1-4 and 9-12, and was 368.6 days in Scenarios 5-8 and 13-16, based on numerical integration.

When extrapolation was required to calculate the RMST, we used a flexible Royston-Parmar parametric splines model with 3 knots (using the RMST R package; https://github.com/scientific-computing-solutions/RMST), consistent with HTA recommendations[30].

### 3.4 Methods to be compared

We compared the following methods:

- *Oracle*: A standard ITT analysis undertaken on the simulated data prior to simulating the impact of switching, which represents the "truth" for each simulation.
- *ITT*: A standard ITT analysis on the simulated data after switching impacted the outcomes.
- *TSE*: Crossover-adjusted analysis using TSE based on only the RCT data with re-censoring applied (Section 2.2).
- *ECA*: A propensity-score based analysis in which the external control data is used as an external control arm (and individuals randomized to control in the RCT are ignored). Specifically, we used the "weightit" function from the WeightIt R library[31] to derive the average treatment effect for the treated (ATT) weights using a logistic regression model. The weighted OS ECA data was then used to estimate the RMST in the control group.

- *ATSE (c=1)*: An analysis based on the proposed approach with decay factor set to *c=1* and re-censoring applied (Section 2.3). This smaller value of *c* will result in borrowing more information from the external data.
- *ATSE (c=4)*: An analysis based on the proposed approach with decay factor set to *c=4* and re-censoring applied (Section 2.3). This method considers a "mid-range" value of *c*.
- *ATSE (c=8)*: An analysis based on the proposed approach with decay factor set to *c=8* and re-censoring applied (Section 2.3). This larger value of *c* will result in borrowing less information from the external data.
- *ATSE (c=4) with adjustment for PFS*: An analysis based on the proposed approach with decay factor set to *c=4* and re-censoring applied (Section 2.3) which includes PFS as a covariate in the ATSE Step 1 and Step 2 AFT models.

### 3.5 Performance measures

Consistent with Latimer et al. (2020)[29], the performance of each method was evaluated according to the percentage bias in the estimate of the control group RMST at study end date. Percentage bias was estimated by taking the difference between the estimated RMST and the true RMST, expressed as a percentage of true RMST. Root mean squared error (RMSE) and empirical standard errors (SE) of the RMST estimates were also calculated for each method and expressed as percentages of the true RMST.

### 3.6 Results

Figure 1 plots the results for Scenario 1 and Figure 2 plots the results for Scenario 9. Figures 3-8 display results in nested loop plots for the percentage bias and empirical SE across all 16 scenarios. Tables 2-5 in the Appendix list the complete results.

Let us focus first on the results from Scenario 1. As expected, under all 3 conditions, the *ITT* analysis over-estimates the control group RMST, resulting in a percentage bias of between 4.4% and 5.4%. Under Condition A, all other methods (*TSE*, *ECA* and *ATSE*) predict the control group RMST with negligible bias (<=0.6%). Amongst these other methods, the *ASTE* analysis appeared to be the most efficient with SEs ranging from 5.4% to 6.2%. Notably, the *ECA* approach had an empirical standard error of 6.3%, lower than the *TSE*, which had an empirical standard error of 7.6%. It is interesting that the SE for the ATSE can be slightly lower than for the Oracle (compare 5.4% to 5.7%). This can be explained by the fact that more data is being used in the ATSE analysis. Under Condition B, when there may be confounding bias with respect to the external control data, the *ECA* appears to be biased, over-estimating the control group RMST by 5.0%. The *ATSE* analyses are also biased, but to a much lesser degree, over-estimating the control group RMST by only 1.2%-1.6%. Under Condition C, when there may be confounding bias with respect to the crossover adjustment, the *TSE* appears to be biased, over-estimating the control group RMST by about 2.2%. The *ATSE* analyses are also biased, over-estimating the control group RMST by about 1.2%-1.5%. Results were similar across all scenarios.

The impact of using different values for the decay factor with *ATSE* was, overall, rather minimal. Recall that a smaller decay factor corresponds to a higher degree of borrowing from the external data. We see that under Condition A, the empirical SE as percentage of the true RMST is smallest when *c*=1 and is highest when *c*=8, meaning that more borrowing leads to more efficiency. Under Condition B, where the external control data introduces bias, the percent bias in RMST is highest when *c*=1, and is lowest when *c*=8. In contrast, under Condition C, the percent bias in RMST is lowest when *c*=1, and is highest when *c*=8. As such, the decay factor seems to correspond to a trade-off between the potential bias due to unmeasured confounding in either the RCT crossover-adjusted controls or the external control.

The impact of adjusting for PFS in the ASTE was notable. We see that the degree of bias due to confounding is substantially reduced at the cost of a modest increase in SE. We do not observe any substantial impact with respect to differing treatment effect sizes ($\delta_1$ = -0.2 vs $\delta_1$ = -0.5), nor do we observe any substantial impact with respect to differing switching proportions or censoring. The ASTE method is consistently the most efficient approach when the sample size of the external control arm was $N_{EC}$=200 (Scenarios 1-8). However, the ECA was notably superior when the sample size of the external control arm was $N_{EC}$=1000 (Scenarios 9-16).

| **Scenario** | **Treatment effect** | **Switch proportion** | **Censoring** | **Sample size of the external control arm** | **True RMST for control group** |
|---|---|---|---|---|---|
| 1 | Low | Moderate | None | 200 | 472.75 |
| 2 | High | Moderate | None | 200 | 472.75 |
| 3 | Low | High | None | 200 | 472.75 |
| 4 | High | High | None | 200 | 472.75 |
| 5 | Low | Moderate | Moderate | 200 | 368.60 |
| 6 | High | Moderate | Moderate | 200 | 368.60 |
| 7 | Low | High | Moderate | 200 | 368.60 |
| 8 | High | High | Moderate | 200 | 368.60 |
| 9 | Low | Moderate | None | 1000 | 472.75 |
| 10 | High | Moderate | None | 1000 | 472.75 |
| 11 | Low | High | None | 1000 | 472.75 |
| 12 | High | High | None | 1000 | 472.75 |
| 13 | Low | Moderate | Moderate | 1000 | 368.60 |
| 14 | High | Moderate | Moderate | 1000 | 368.60 |
| 15 | Low | High | Moderate | 1000 | 368.60 |
| 16 | High | High | Moderate | 1000 | 368.60 |

**Table 1. Overview of simulated scenarios**

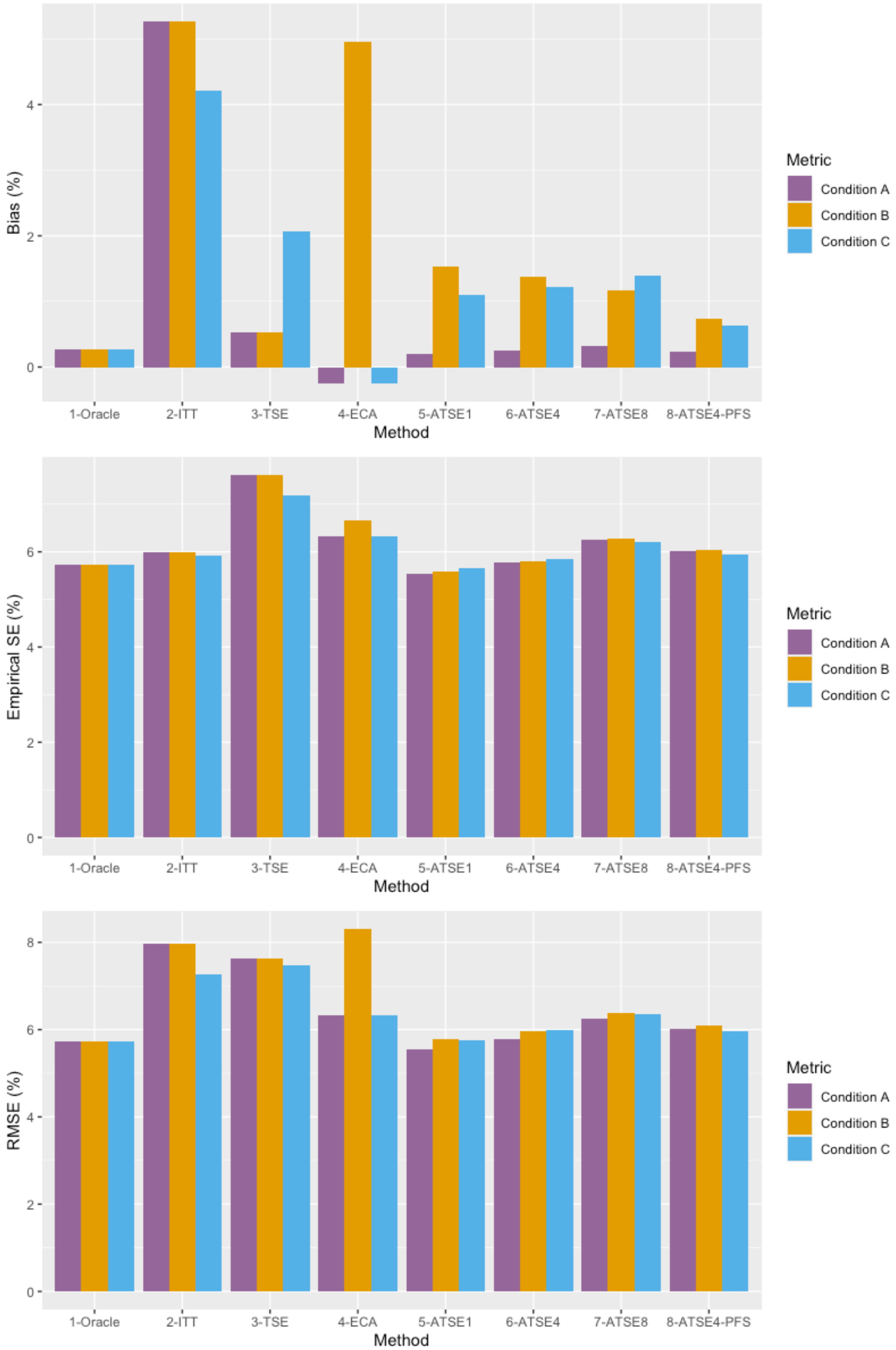

**Figure 2 – Simulation study results from Scenario 1.**

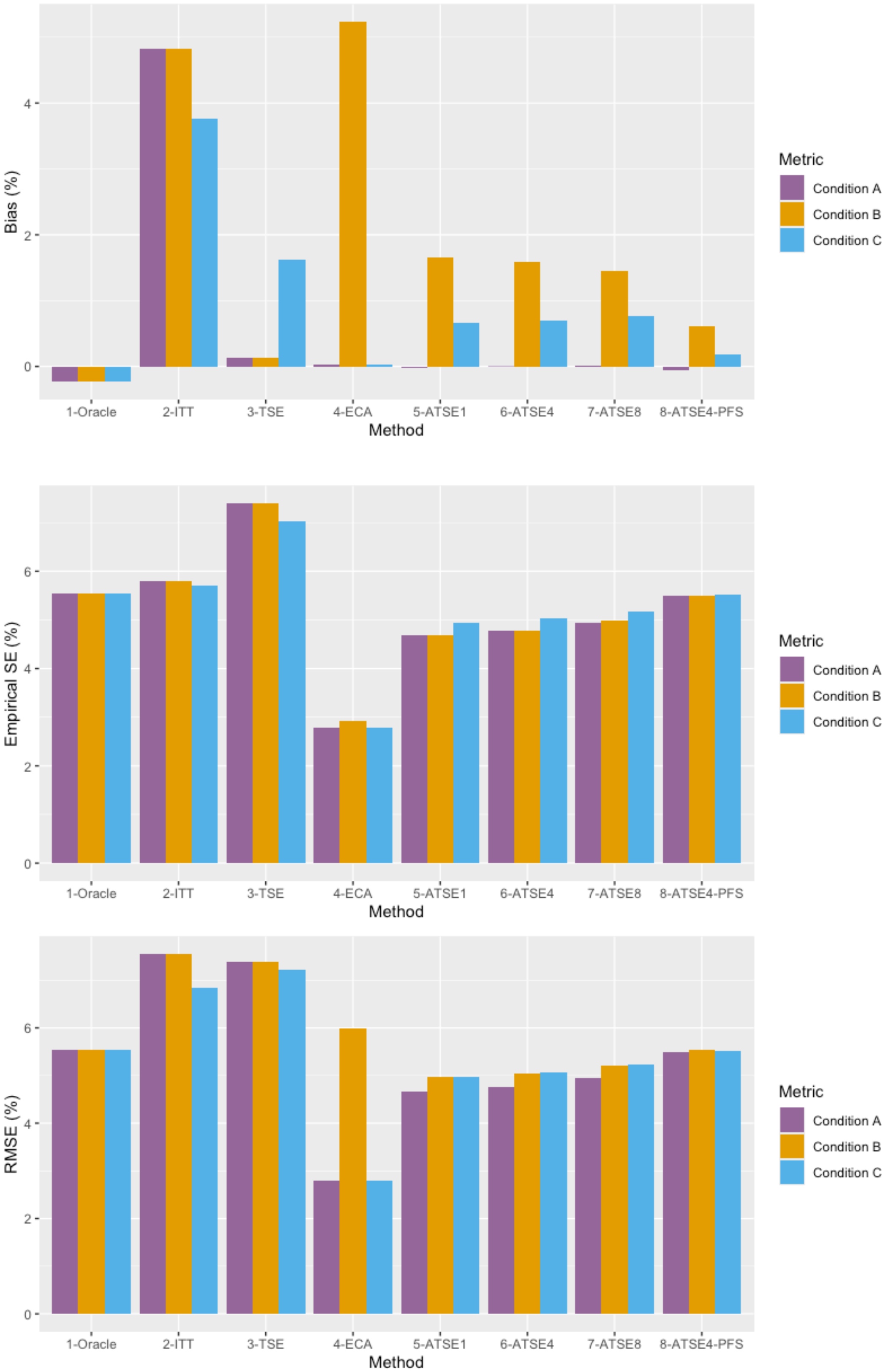

**Figure 3 – Simulation study results from Scenario 9.**

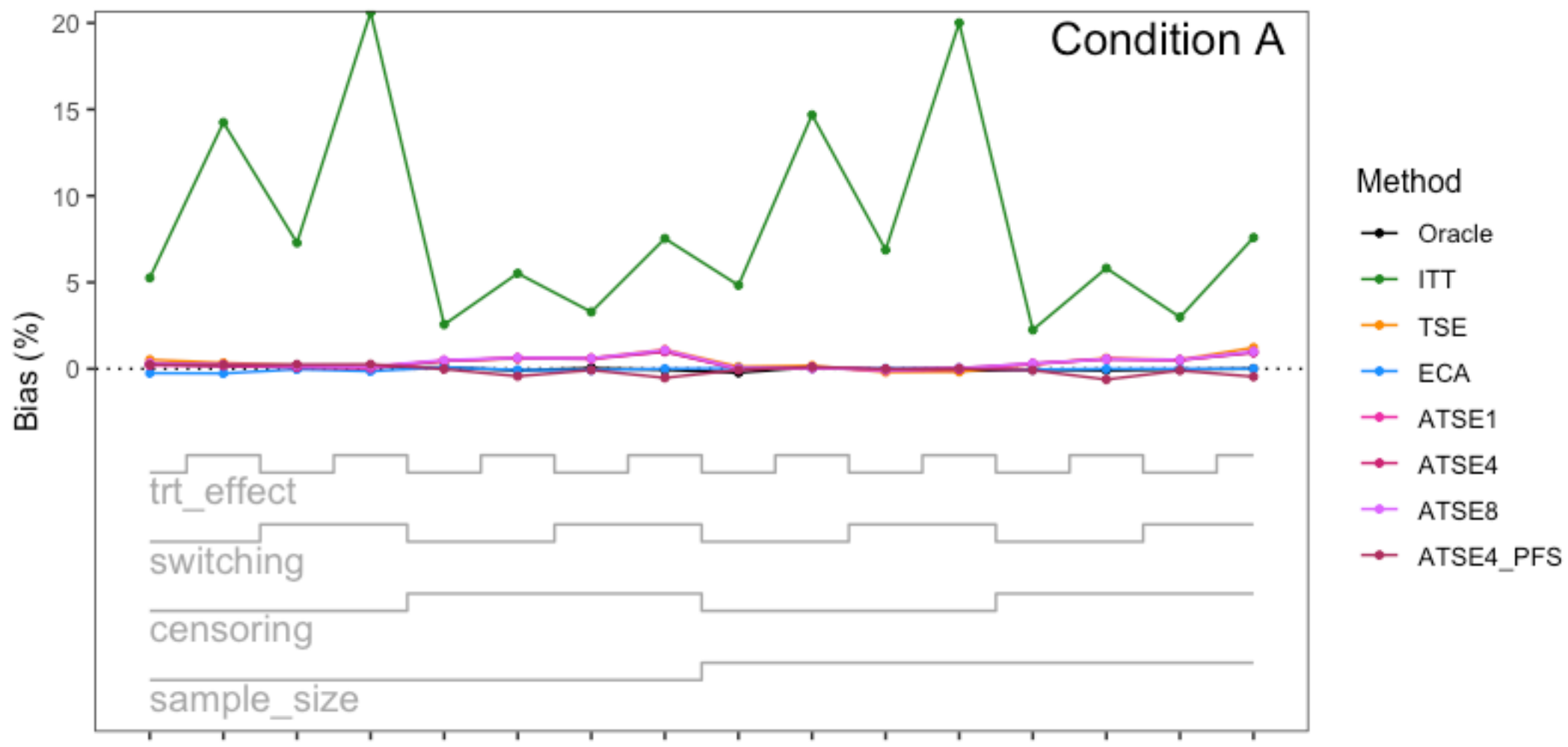

**Figure 4 – Simulation study results: Percentage bias for Condition A across the 16 scenarios.**

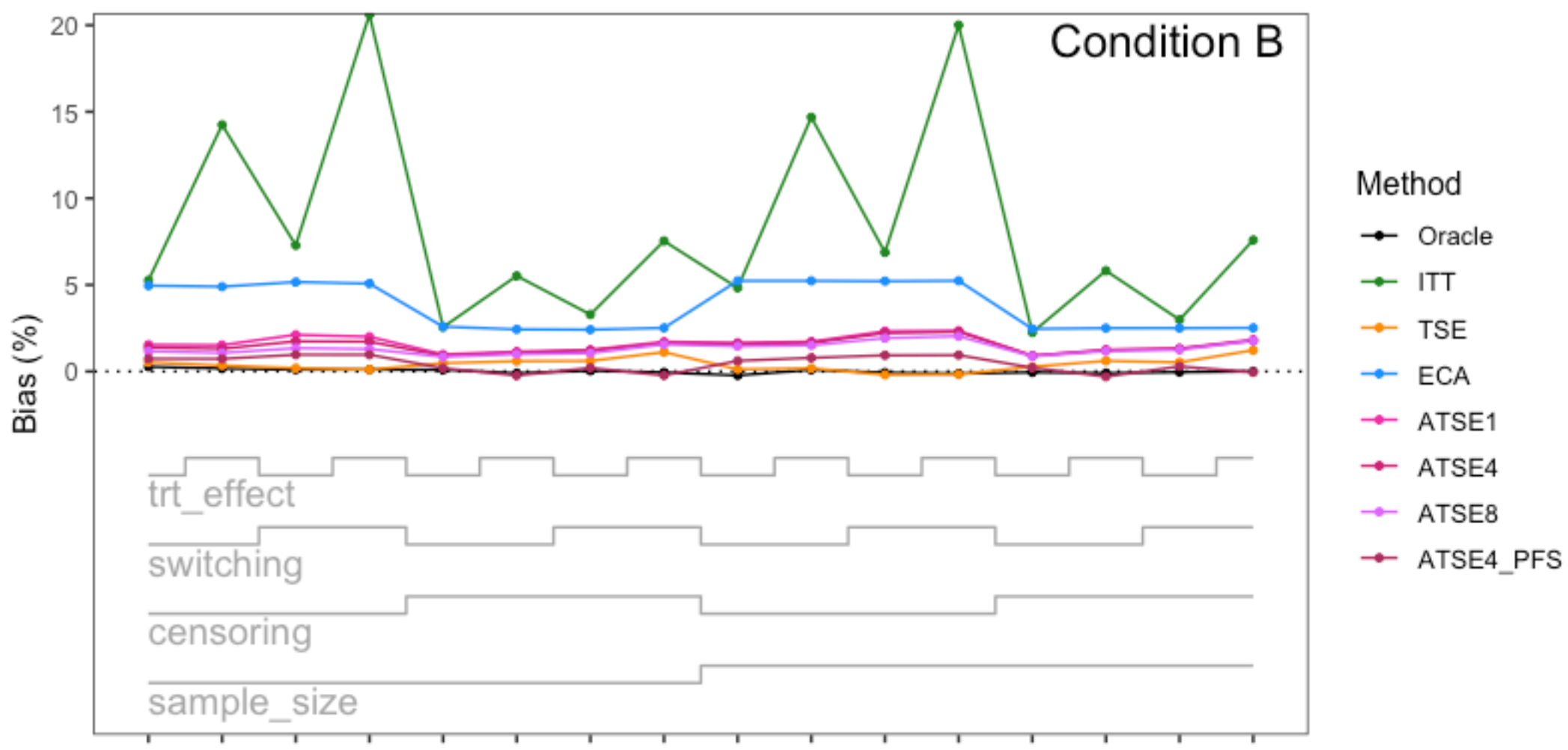

**Figure 5 – Simulation study results: Percentage bias for Condition B across the 16 scenarios.**

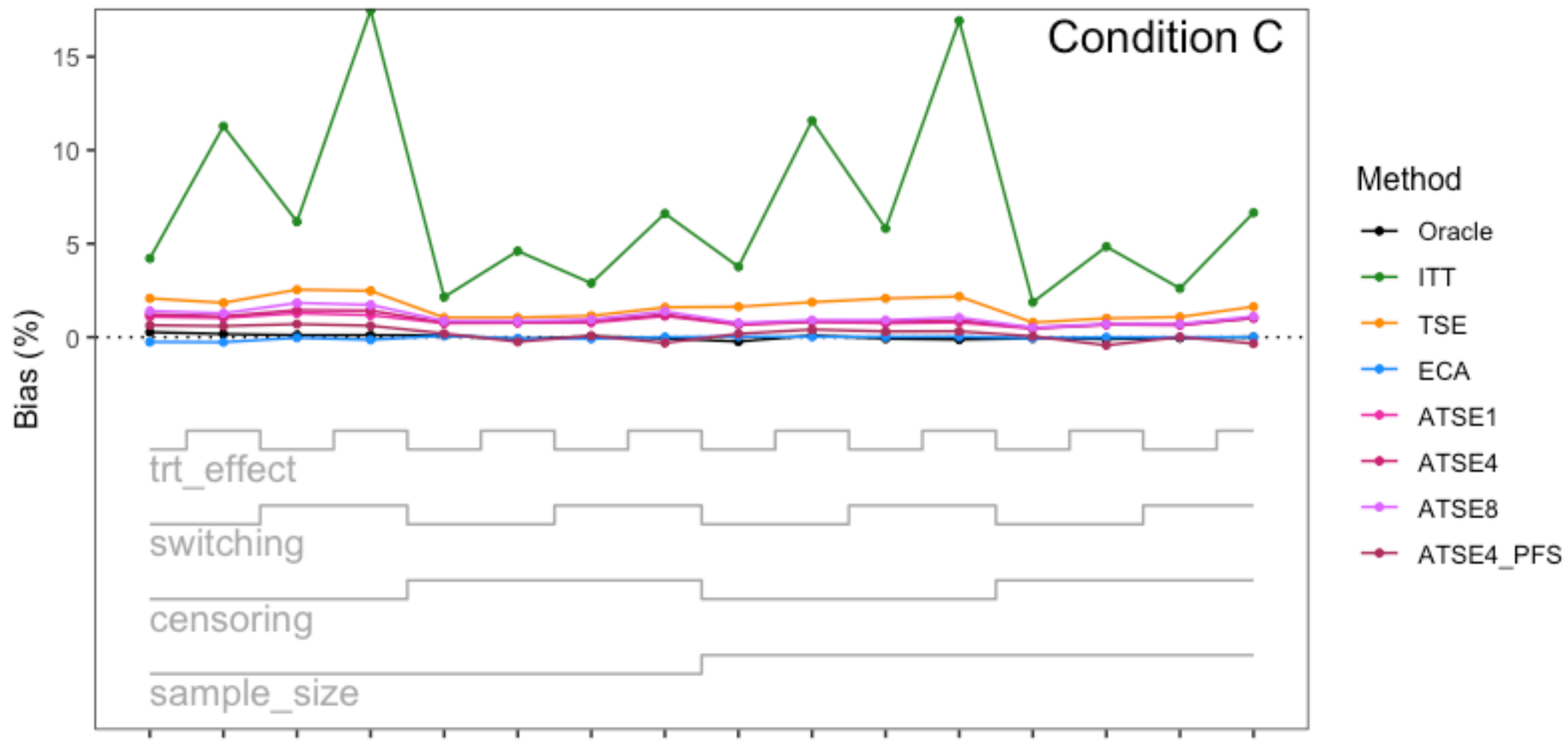

**Figure 6 – Simulation study results: Percentage bias for Condition C across the 16 scenarios.**

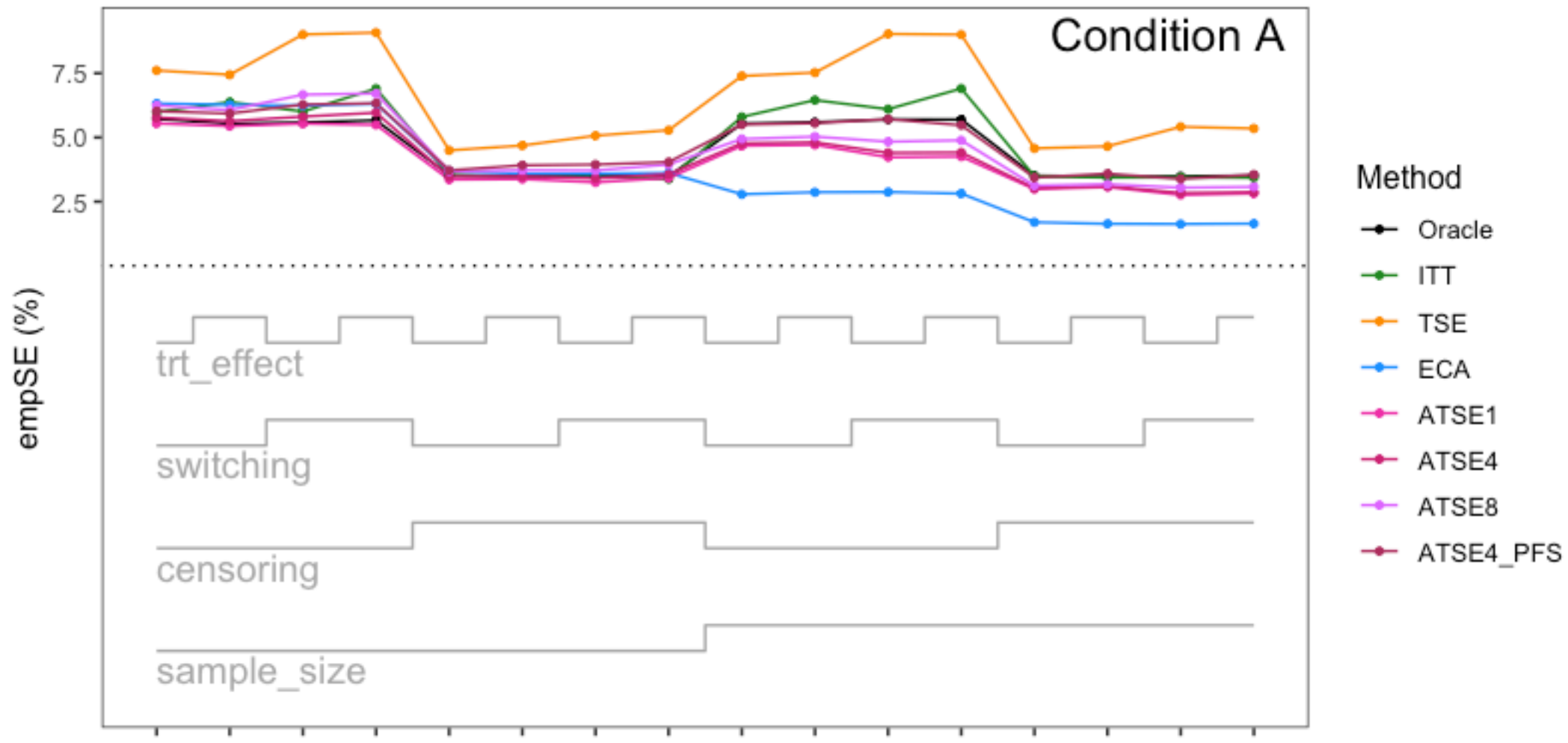

**Figure 7 – Simulation study results: Empirical SE for Condition A across the 16 scenarios.**

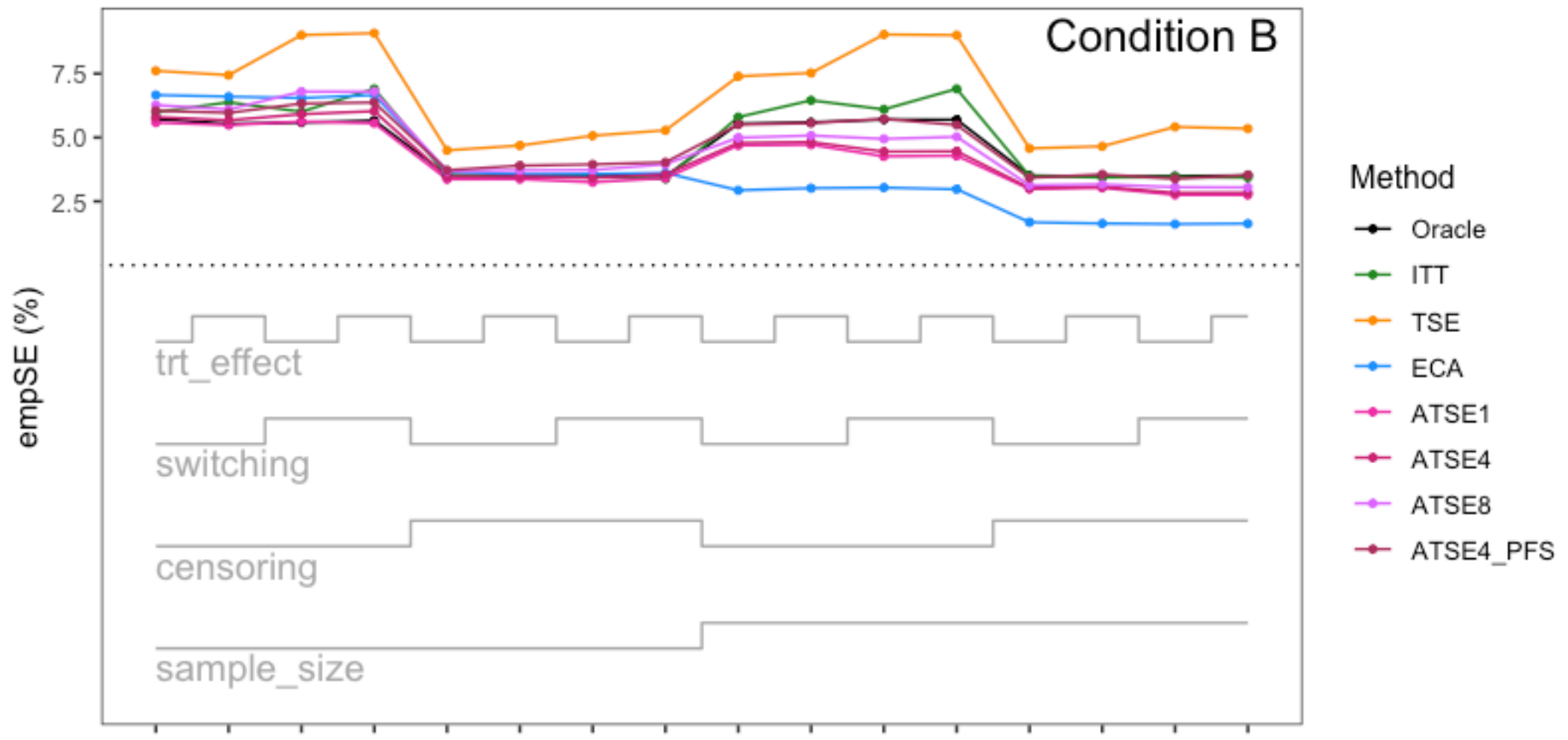

**Figure 8 – Simulation study results: Empirical SE for Condition B across the 16 scenarios.**

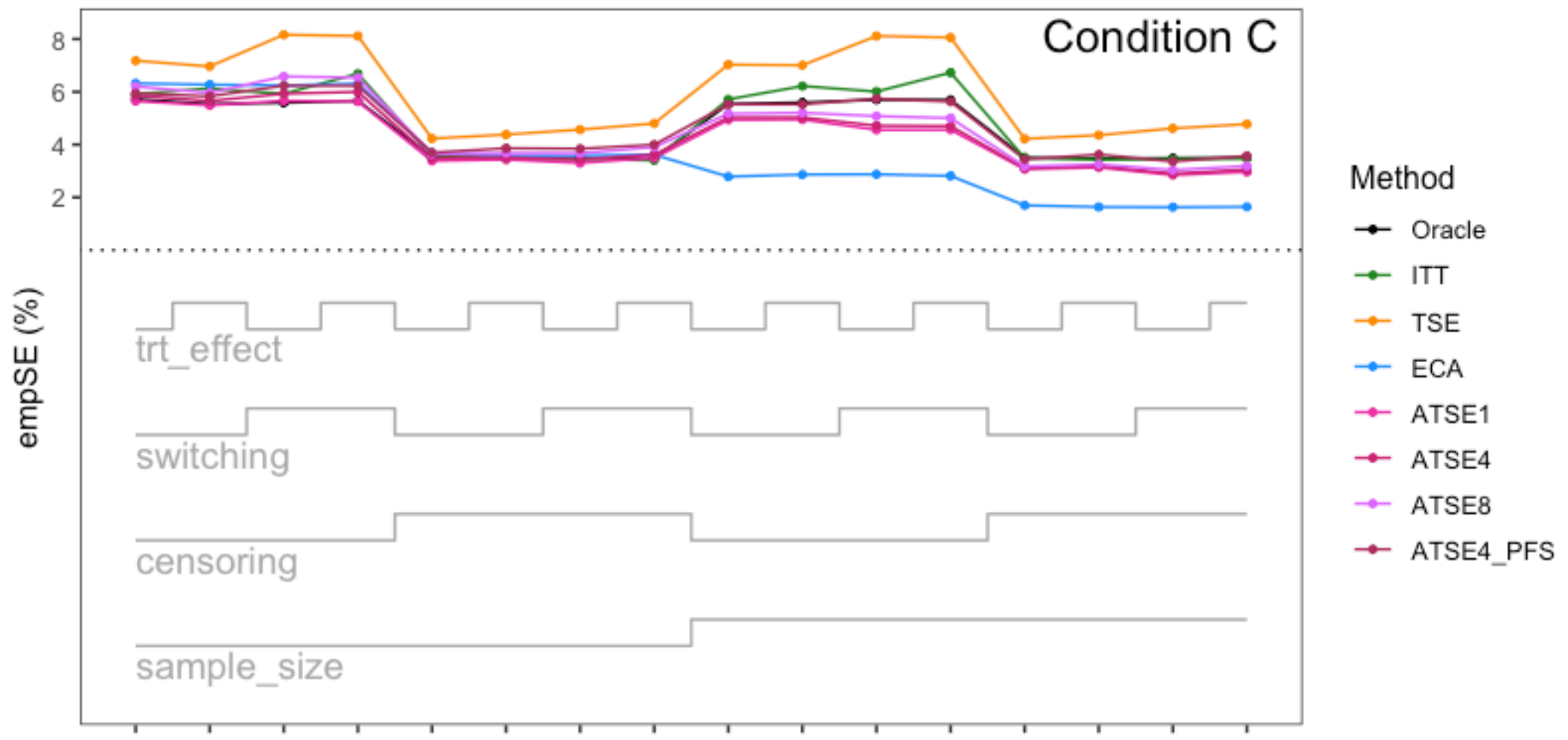

**Figure 9 – Simulation study results: Empirical SE for Condition C across the 16 scenarios.**

## 4 Conclusion

The TSD 24 specifically brings attention to the potential for "external data [is used] to estimate counterfactual survival beyond the switching time-point" and suggests that "further research [to develop such methods] may be valuable" (TSD 24, April, 2024). In this paper, we proposed a new approach, the ATSE, which may indeed be prove valuable.

While RWE can be used as an external control arm to completely replace a crossover contaminated RCT control arm, this strategy discards a large amount of potentially valuable data (including the PFS times of those RCT subjects randomized to control) and can be susceptible to bias due to unmeasured confounding[32]. As illustrated in the simulation study, the ATSE approach leverages all the available data (and only the necessary external data) and consequently may be less impacted by confounding bias. This aligns with the current understanding that hybrid control arm studies (also known as "augmented RCTs") should be considered a higher level of evidence than external control arm studies; see Gray et al. (2020)[33].

If the assumptions underlying the crossover adjustment are suspect or the size of the external dataset is very large, then the ECA may be preferable to the ATSE approach. Alternatively, if the assumption of exchangeability for the subjects in the external data is suspect, then the TSE may be preferable to the ATSE approach. Figure 10 illustrates where the different methods might fit in terms of deciding upon the most appropriate approach.

When compared to the standard TSE approach, the ATSE approach has potential to obtain more precise estimation of survival outcomes particularly when sample sizes are small, and crossover rates are high. However, the need for strong assumptions remains. With TSE, the assumption of no unmeasured confounding implies that, conditional on all the observed covariates, a participant's decision to switch is independent of their post-progression survival. With ATSE, an additional assumption is also required: Conditional on all observed covariates, individuals from the RCT and the external cohort must be exchangeable; see Bours (2020)[34]. The ATSE method also requires one to pre-specify a value for the decay factor which will impact the overall amount of borrowing. Tan et al. (2022) recommend that this value be determined prior to analyzing the data by means of a simulation study and based Tan et al. (2022)'s simulation study, Sengupta et al. (2023)[18] use a value of $c = 4$ for their case study. In our simulation study, we considered $c = 1, 4, \text{and } 8$ and found that the results did not differ substantially across the three different values. Following one's analysis, the amount of borrowing can be assessed using the "effective number of external events", see Sengupta et al. (2023)[18].

While the ATSE method represents a promising avenue for addressing treatment switching in clinical trials, its limitations must be acknowledged and carefully considered. One notable limitation is the assumption that treatment switching occurs exclusively at, or shortly after, disease progression. This assumption may not always hold true in real-world scenarios, where switching can occur for various reasons and at different points in time. Our simulation study was simply intended to demonstrate a proof of concept in simple scenarios, and as such we did not investigate the impact of time-dependent confounding. The g-computation TSE method, proposed by see Latimer et al. (2020)[29], offers a potential solution for situations where switching is not confined to the point of progression (or to another well-defined "secondary baseline"). Further research is needed to explore the feasibility of augmenting the g-computation TSE with external data.

Another potential limitation of ASTE might be the need for comprehensive data on all confounding variables. To avoid any bias due to unmeasured confounding one must adjust for any factors that could simultaneously influence survival, switching and participation in the RCT vs external cohorts. Identifying these factors could prove difficult and the conservative approach of simply adjusting for all variables that could be considered important prognostic factors requires substantial data availability. The dynamic borrowing approach we propose within the ATSE method can help reduce the amount of bias due to unmeasured confounding but cannot entirely eliminate it. Adjusting for PFS as a "proxy confounder" might help further mitigate the risk of bias due to unmeasured confounding.

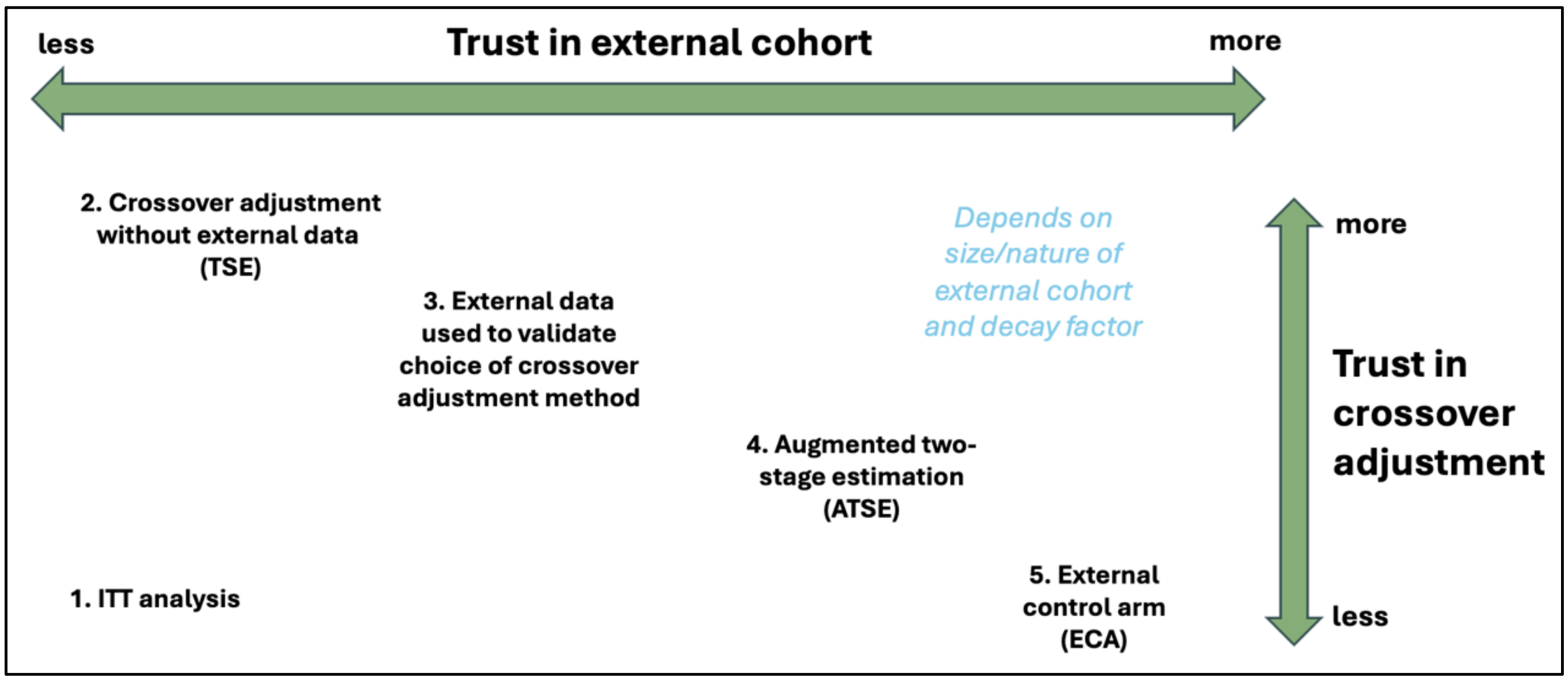

**Figure 10.** If the assumptions underlying the crossover adjustment are suspect, then the ECA may be preferable to the ATSE approach. Alternatively, if the assumption of exchangeability for the subjects in the external data is suspect, then the TSE may be preferable to the ATSE approach. If no assumptions can be relied upon, the ITT analysis (while biased) will be most appropriate.

## Appendix

| | Method | Percent bias in RMST | | | Empirical SE as percentage of true RMST | | | RMSE as percentage of true RMST | | |
|---|---|---|---|---|---|---|---|---|---|---|
| **Condition** | | **A** | **B** | **C** | **A** | **B** | **C** | **A** | **B** | **C** |
| **Scenario 1** | Oracle | 0.30 | 0.30 | 0.30 | 5.70 | 5.70 | 5.70 | 5.70 | 5.70 | 5.70 |
| *Low* | ITT | 5.30 | 5.30 | 4.20 | 6.00 | 6.00 | 5.90 | 8.00 | 8.00 | 7.30 |
| *treatment* | TSE | 0.50 | 0.50 | 2.10 | 7.60 | 7.60 | 7.20 | 7.60 | 7.60 | 7.50 |
| *effect &* | ECA | -0.30 | 5.00 | -0.30 | 6.30 | 6.70 | 6.30 | 6.30 | 8.30 | 6.30 |
| *Moderate* | ATSE (*c=1*) | 0.20 | 1.50 | 1.10 | 5.50 | 5.60 | 5.60 | 5.50 | 5.80 | 5.80 |
| *switching* | ATSE (*c=4*) | 0.20 | 1.40 | 1.20 | 5.80 | 5.80 | 5.90 | 5.80 | 6.00 | 6.00 |
| *proportion,* | ATSE (*c=8*) | 0.30 | 1.20 | 1.40 | 6.20 | 6.30 | 6.20 | 6.30 | 6.40 | 6.40 |
| $N_{EC}$=200 | ATSE (*c=4*) adj. for PFS | 0.20 | 0.70 | 0.60 | 6.00 | 6.00 | 5.90 | 6.00 | 6.10 | 6.00 |
| **Scenario 2** | Oracle | 0.17 | 0.17 | 0.17 | 5.54 | 5.54 | 5.54 | 5.54 | 5.54 | 5.54 |
| *High* | ITT | 14.24 | 14.24 | 11.28 | 6.38 | 6.38 | 6.13 | 15.61 | 15.61 | 12.84 |
| *treatment* | TSE | 0.34 | 0.34 | 1.84 | 7.44 | 7.44 | 6.96 | 7.45 | 7.45 | 7.20 |
| *effect &* | ECA | -0.27 | 4.90 | -0.27 | 6.27 | 6.60 | 6.27 | 6.28 | 8.22 | 6.28 |
| *Moderate* | ATSE (*c=1*) | 0.18 | 1.52 | 1.03 | 5.44 | 5.49 | 5.49 | 5.44 | 5.69 | 5.58 |
| *switching* | ATSE (*c=4*) | 0.18 | 1.31 | 1.11 | 5.63 | 5.67 | 5.66 | 5.63 | 5.82 | 5.76 |
| *proportion,* | ATSE (*c=8*) | 0.25 | 1.07 | 1.28 | 6.07 | 6.10 | 5.96 | 6.07 | 6.19 | 6.09 |
| $N_{EC}$=200 | ATSE (*c=4*) adj. for PFS | 0.24 | 0.73 | 0.60 | 5.93 | 5.96 | 5.83 | 5.93 | 6.01 | 5.86 |
| **Scenario 3** | Oracle | 0.11 | 0.11 | 0.11 | 5.58 | 5.58 | 5.58 | 5.58 | 5.58 | 5.58 |
| *Low* | ITT | 7.28 | 7.28 | 6.18 | 6.01 | 6.01 | 5.93 | 9.44 | 9.44 | 8.56 |
| *treatment* | TSE | 0.19 | 0.19 | 2.54 | 9.01 | 9.01 | 8.16 | 9.01 | 9.01 | 8.55 |
| *effect &* | ECA | -0.03 | 5.16 | -0.03 | 6.24 | 6.54 | 6.24 | 6.24 | 8.33 | 6.24 |
| *High* | ATSE (*c=1*) | 0.14 | 2.12 | 1.29 | 5.54 | 5.62 | 5.66 | 5.54 | 6.01 | 5.80 |
| *switching* | ATSE (*c=4*) | 0.08 | 1.74 | 1.45 | 5.81 | 5.91 | 5.93 | 5.81 | 6.16 | 6.10 |
| *proportion,* | ATSE (*c=8*) | 0.17 | 1.34 | 1.83 | 6.66 | 6.79 | 6.58 | 6.67 | 6.92 | 6.83 |
| $N_{EC}$=200 | ATSE (*c=4*) adj. for PFS | 0.25 | 0.96 | 0.70 | 6.28 | 6.33 | 6.23 | 6.28 | 6.40 | 6.27 |
| **Scenario 4** | Oracle | 0.11 | 0.11 | 0.11 | 5.66 | 5.66 | 5.66 | 5.66 | 5.66 | 5.66 |
| *High* | ITT | 20.61 | 20.61 | 17.48 | 6.90 | 6.90 | 6.69 | 21.73 | 21.73 | 18.71 |
| *treatment* | TSE | 0.10 | 0.10 | 2.48 | 9.08 | 9.08 | 8.12 | 9.08 | 9.08 | 8.48 |
| *effect &* | ECA | -0.14 | 5.07 | -0.14 | 6.31 | 6.66 | 6.31 | 6.31 | 8.37 | 6.31 |
| *High* | ATSE (*c=1*) | 0.03 | 2.00 | 1.18 | 5.49 | 5.56 | 5.64 | 5.49 | 5.91 | 5.76 |
| *switching* | ATSE (*c=4*) | 0.08 | 1.72 | 1.42 | 5.96 | 6.03 | 6.00 | 5.96 | 6.27 | 6.16 |
| *proportion,* | ATSE (*c=8*) | 0.13 | 1.29 | 1.73 | 6.71 | 6.79 | 6.54 | 6.71 | 6.91 | 6.76 |
| $N_{EC}$=200 | ATSE (*c=4*) adj. for PFS | 0.26 | 0.96 | 0.62 | 6.33 | 6.37 | 6.23 | 6.33 | 6.44 | 6.26 |

**Table 2. Simulation study results for Scenarios 1-4 (no censoring; small external control, $N_{EC}$=200).**

| | Method | Percent bias in RMST | | | Empirical SE as percentage of true RMST | | | RMSE as percentage of true RMST | | |
|---|---|---|---|---|---|---|---|---|---|---|
| **Condition** | | **A** | **B** | **C** | **A** | **B** | **C** | **A** | **B** | **C** |
| **Scenario 5** | Oracle | 0.09 | 0.09 | 0.09 | 3.53 | 3.53 | 3.53 | 3.53 | 3.53 | 3.53 |
| *Low* | ITT | 2.56 | 2.56 | 2.15 | 3.53 | 3.53 | 3.53 | 4.35 | 4.35 | 4.13 |
| *treatment* | TSE | 0.48 | 0.48 | 1.05 | 4.50 | 4.50 | 4.23 | 4.52 | 4.52 | 4.36 |
| *effect &* | ECA | 0.08 | 2.59 | 0.08 | 3.69 | 3.66 | 3.69 | 3.69 | 4.48 | 3.69 |
| *Moderate* | ATSE (*c=1*) | 0.47 | 0.98 | 0.75 | 3.36 | 3.37 | 3.38 | 3.39 | 3.51 | 3.47 |
| *switching* | ATSE (*c=4*) | 0.46 | 0.91 | 0.78 | 3.47 | 3.46 | 3.48 | 3.50 | 3.58 | 3.57 |
| *proportion,* | ATSE (*c=8*) | 0.51 | 0.84 | 0.87 | 3.68 | 3.68 | 3.65 | 3.71 | 3.78 | 3.75 |
| $N_{EC}$=200 | ATSE (*c=4*) adj. for PFS | -0.02 | 0.19 | 0.20 | 3.72 | 3.72 | 3.70 | 3.72 | 3.72 | 3.70 |
| **Scenario 6** | Oracle | -0.10 | -0.10 | -0.10 | 3.50 | 3.50 | 3.50 | 3.50 | 3.50 | 3.50 |
| *High* | ITT | 5.52 | 5.52 | 4.61 | 3.45 | 3.45 | 3.46 | 6.51 | 6.51 | 5.76 |
| *treatment* | TSE | 0.59 | 0.59 | 1.05 | 4.68 | 4.68 | 4.38 | 4.72 | 4.72 | 4.50 |
| *effect &* | ECA | -0.07 | 2.43 | -0.07 | 3.59 | 3.56 | 3.59 | 3.59 | 4.31 | 3.59 |
| *Moderate* | ATSE (*c=1*) | 0.62 | 1.14 | 0.77 | 3.38 | 3.37 | 3.43 | 3.43 | 3.56 | 3.52 |
| *switching* | ATSE (*c=4*) | 0.62 | 1.07 | 0.81 | 3.48 | 3.47 | 3.50 | 3.53 | 3.63 | 3.59 |
| *proportion,* | ATSE (*c=8*) | 0.63 | 1.00 | 0.86 | 3.72 | 3.71 | 3.68 | 3.78 | 3.84 | 3.78 |
| $N_{EC}$=200 | ATSE (*c=4*) adj. for PFS | -0.43 | -0.24 | -0.25 | 3.91 | 3.90 | 3.86 | 3.94 | 3.90 | 3.87 |
| **Scenario 7** | Oracle | 0.04 | 0.04 | 0.04 | 3.49 | 3.49 | 3.49 | 3.49 | 3.49 | 3.49 |
| *Low* | ITT | 3.29 | 3.29 | 2.89 | 3.46 | 3.46 | 3.46 | 4.77 | 4.77 | 4.51 |
| *treatment* | TSE | 0.61 | 0.61 | 1.14 | 5.07 | 5.07 | 4.57 | 5.10 | 5.10 | 4.71 |
| *effect &* | ECA | -0.08 | 2.41 | -0.08 | 3.59 | 3.57 | 3.59 | 3.59 | 4.31 | 3.59 |
| *High* | ATSE (*c=1*) | 0.56 | 1.24 | 0.78 | 3.25 | 3.25 | 3.30 | 3.30 | 3.48 | 3.39 |
| *switching* | ATSE (*c=4*) | 0.60 | 1.20 | 0.86 | 3.41 | 3.41 | 3.44 | 3.46 | 3.62 | 3.54 |
| *proportion,* | ATSE (*c=8*) | 0.61 | 1.05 | 0.94 | 3.71 | 3.73 | 3.67 | 3.76 | 3.87 | 3.79 |
| $N_{EC}$=200 | ATSE (*c=4*) adj. for PFS | -0.07 | 0.20 | 0.08 | 3.94 | 3.95 | 3.84 | 3.94 | 3.95 | 3.84 |
| **Scenario 8** | Oracle | -0.06 | -0.06 | -0.06 | 3.48 | 3.48 | 3.48 | 3.48 | 3.48 | 3.48 |
| *High* | ITT | 7.53 | 7.53 | 6.61 | 3.38 | 3.38 | 3.40 | 8.25 | 8.25 | 7.44 |
| *treatment* | TSE | 1.11 | 1.11 | 1.60 | 5.28 | 5.28 | 4.80 | 5.39 | 5.39 | 5.06 |
| *effect &* | ECA | 0.00 | 2.51 | 0.00 | 3.62 | 3.61 | 3.62 | 3.61 | 4.40 | 3.61 |
| *High* | ATSE (*c=1*) | 0.99 | 1.71 | 1.13 | 3.42 | 3.41 | 3.50 | 3.56 | 3.81 | 3.68 |
| *switching* | ATSE (*c=4*) | 1.00 | 1.63 | 1.21 | 3.56 | 3.55 | 3.62 | 3.70 | 3.91 | 3.81 |
| *proportion,* | ATSE (*c=8*) | 1.09 | 1.56 | 1.38 | 3.95 | 3.95 | 3.91 | 4.09 | 4.25 | 4.14 |
| $N_{EC}$=200 | ATSE (*c=4*) adj. for PFS | -0.52 | -0.23 | -0.30 | 4.04 | 4.03 | 4.00 | 4.07 | 4.03 | 4.01 |

**Table 3. Simulation study results for Scenarios 5-8 (moderate censoring; small external control, $N_{EC}$=200).**

| | Method | Percent bias in RMST | | | Empirical SE as percentage of true RMST | | | RMSE as percentage of true RMST | | |
|---|---|---|---|---|---|---|---|---|---|---|
| **Condition** | | **A** | **B** | **C** | **A** | **B** | **C** | **A** | **B** | **C** |
| **Scenario 9** | Oracle | -0.23 | -0.23 | -0.23 | 5.54 | 5.54 | 5.54 | 5.54 | 5.54 | 5.54 |
| *Low treatment* | ITT | 4.83 | 4.83 | 3.77 | 5.80 | 5.80 | 5.71 | 7.54 | 7.54 | 6.84 |
| *effect &* | TSE | 0.12 | 0.12 | 1.63 | 7.39 | 7.39 | 7.03 | 7.39 | 7.39 | 7.22 |
| *Moderate* | ECA | 0.04 | 5.23 | 0.04 | 2.79 | 2.93 | 2.79 | 2.79 | 6.00 | 2.79 |
| *switching* | ATSE (*c=1*) | -0.02 | 1.65 | 0.67 | 4.67 | 4.69 | 4.94 | 4.67 | 4.97 | 4.98 |
| *proportion,* | ATSE (*c=4*) | -0.01 | 1.59 | 0.70 | 4.77 | 4.79 | 5.03 | 4.77 | 5.04 | 5.07 |
| $N_{EC}$=1000 | ATSE (*c=8*) | 0.01 | 1.46 | 0.76 | 4.94 | 4.99 | 5.17 | 4.94 | 5.20 | 5.23 |
| | ATSE (*c=4*) adj. for PFS | -0.06 | 0.61 | 0.18 | 5.50 | 5.50 | 5.53 | 5.50 | 5.54 | 5.53 |
| **Scenario 10** | Oracle | 0.09 | 0.09 | 0.09 | 5.59 | 5.59 | 5.59 | 5.59 | 5.59 | 5.59 |
| *High treatment* | ITT | 14.68 | 14.68 | 11.58 | 6.45 | 6.45 | 6.22 | 16.04 | 16.04 | 13.14 |
| *effect &* | TSE | 0.18 | 0.18 | 1.87 | 7.52 | 7.52 | 7.01 | 7.53 | 7.53 | 7.25 |
| *Moderate* | ECA | 0.02 | 5.23 | 0.02 | 2.86 | 3.02 | 2.86 | 2.86 | 6.04 | 2.86 |
| *switching* | ATSE (*c=1*) | 0.04 | 1.72 | 0.82 | 4.71 | 4.72 | 4.95 | 4.71 | 5.02 | 5.02 |
| *proportion,* | ATSE (*c=4*) | 0.05 | 1.65 | 0.85 | 4.81 | 4.82 | 5.03 | 4.81 | 5.09 | 5.10 |
| $N_{EC}$=1000 | ATSE (*c=8*) | 0.06 | 1.51 | 0.91 | 5.03 | 5.08 | 5.20 | 5.03 | 5.29 | 5.28 |
| | ATSE (*c=4*) adj. for PFS | 0.11 | 0.79 | 0.41 | 5.56 | 5.56 | 5.52 | 5.56 | 5.62 | 5.54 |
| **Scenario 11** | Oracle | -0.07 | -0.07 | -0.07 | 5.70 | 5.70 | 5.70 | 5.70 | 5.70 | 5.70 |
| *Low treatment* | ITT | 6.88 | 6.88 | 5.81 | 6.10 | 6.10 | 6.01 | 9.19 | 9.19 | 8.36 |
| *effect & High* | TSE | -0.20 | -0.20 | 2.07 | 9.03 | 9.03 | 8.11 | 9.03 | 9.03 | 8.37 |
| *switching* | ECA | 0.02 | 5.21 | 0.02 | 2.87 | 3.04 | 2.87 | 2.87 | 6.03 | 2.87 |
| *proportion,* | ATSE (*c=1*) | -0.06 | 2.32 | 0.75 | 4.23 | 4.26 | 4.56 | 4.23 | 4.85 | 4.62 |
| $N_{EC}$=1000 | ATSE (*c=4*) | -0.04 | 2.22 | 0.83 | 4.40 | 4.45 | 4.72 | 4.40 | 4.97 | 4.79 |
| | ATSE (*c=8*) | -0.07 | 1.92 | 0.91 | 4.83 | 4.95 | 5.08 | 4.83 | 5.31 | 5.16 |
| | ATSE (*c=4*) adj. for PFS | -0.01 | 0.93 | 0.31 | 5.71 | 5.73 | 5.74 | 5.71 | 5.80 | 5.75 |
| **Scenario 12** | Oracle | -0.13 | -0.13 | -0.13 | 5.70 | 5.70 | 5.70 | 5.70 | 5.70 | 5.70 |
| *High treatment* | ITT | 20.00 | 20.00 | 16.92 | 6.90 | 6.90 | 6.73 | 21.16 | 21.16 | 18.21 |
| *effect & High* | TSE | -0.18 | -0.18 | 2.18 | 9.00 | 9.00 | 8.06 | 9.00 | 9.00 | 8.35 |
| *switching* | ECA | 0.04 | 5.24 | 0.04 | 2.81 | 2.97 | 2.81 | 2.81 | 6.02 | 2.81 |
| *proportion,* | ATSE (*c=1*) | -0.03 | 2.35 | 0.79 | 4.25 | 4.28 | 4.56 | 4.25 | 4.89 | 4.63 |
| $N_{EC}$=1000 | ATSE (*c=4*) | 0.03 | 2.29 | 0.89 | 4.41 | 4.46 | 4.70 | 4.41 | 5.01 | 4.78 |
| | ATSE (*c=8*) | 0.03 | 2.03 | 1.05 | 4.89 | 5.02 | 5.00 | 4.89 | 5.42 | 5.11 |
| | ATSE (*c=4*) adj. for PFS | 0.01 | 0.94 | 0.32 | 5.47 | 5.49 | 5.63 | 5.47 | 5.57 | 5.64 |

**Table 4. Simulation study results for Scenarios 9-12 (no censoring; large external control, $N_{EC}$=1000).**

| | Method | Percent bias in RMST | | | Empirical SE as percentage of true RMST | | | RMSE as percentage of true RMST | | |
|---|---|---|---|---|---|---|---|---|---|---|
| **Condition** | | **A** | **B** | **C** | **A** | **B** | **C** | **A** | **B** | **C** |
| **Scenario 13** *Low treatment effect & Moderate switching proportion,* $N_{EC}$=1000 | Oracle | -0.05 | -0.05 | -0.05 | 3.51 | 3.51 | 3.51 | 3.51 | 3.51 | 3.51 |
| | ITT | 2.25 | 2.25 | 1.87 | 3.49 | 3.49 | 3.48 | 4.15 | 4.15 | 3.95 |
| | TSE | 0.27 | 0.27 | 0.79 | 4.57 | 4.57 | 4.22 | 4.58 | 4.58 | 4.29 |
| | ECA | -0.05 | 2.46 | -0.05 | 1.70 | 1.68 | 1.70 | 1.70 | 2.98 | 1.70 |
| | ATSE (*c=1*) | 0.26 | 0.89 | 0.46 | 2.99 | 2.99 | 3.07 | 3.00 | 3.12 | 3.10 |
| | ATSE (*c=4*) | 0.31 | 0.91 | 0.50 | 3.03 | 3.02 | 3.10 | 3.04 | 3.15 | 3.14 |
| | ATSE (*c=8*) | 0.31 | 0.87 | 0.51 | 3.12 | 3.13 | 3.17 | 3.14 | 3.25 | 3.21 |
| | ATSE (*c=4*) adj. for PFS | -0.08 | 0.22 | 0.05 | 3.44 | 3.42 | 3.44 | 3.44 | 3.42 | 3.44 |
| **Scenario 14** *High treatment effect & Moderate switching proportion,* $N_{EC}$=1000 | Oracle | -0.10 | -0.10 | -0.10 | 3.47 | 3.47 | 3.47 | 3.47 | 3.47 | 3.47 |
| | ITT | 5.82 | 5.82 | 4.84 | 3.44 | 3.44 | 3.43 | 6.76 | 6.76 | 5.93 |
| | TSE | 0.61 | 0.61 | 1.01 | 4.65 | 4.65 | 4.36 | 4.69 | 4.69 | 4.47 |
| | ECA | 0.00 | 2.50 | 0.00 | 1.64 | 1.63 | 1.64 | 1.64 | 2.99 | 1.64 |
| | ATSE (*c=1*) | 0.57 | 1.25 | 0.68 | 3.07 | 3.03 | 3.13 | 3.12 | 3.28 | 3.21 |
| | ATSE (*c=4*) | 0.55 | 1.22 | 0.68 | 3.10 | 3.08 | 3.19 | 3.15 | 3.31 | 3.26 |
| | ATSE (*c=8*) | 0.56 | 1.18 | 0.71 | 3.18 | 3.16 | 3.25 | 3.22 | 3.38 | 3.33 |
| | ATSE (*c=4*) adj. for PFS | -0.62 | -0.30 | -0.43 | 3.59 | 3.56 | 3.63 | 3.64 | 3.57 | 3.65 |
| **Scenario 15** *Low treatment effect & High switching proportion,* $N_{EC}$=1000 | Oracle | -0.04 | -0.04 | -0.04 | 3.49 | 3.49 | 3.49 | 3.49 | 3.49 | 3.49 |
| | ITT | 2.99 | 2.99 | 2.61 | 3.44 | 3.44 | 3.45 | 4.56 | 4.56 | 4.32 |
| | TSE | 0.52 | 0.52 | 1.08 | 5.41 | 5.41 | 4.62 | 5.44 | 5.44 | 4.74 |
| | ECA | 0.00 | 2.50 | 0.00 | 1.63 | 1.61 | 1.63 | 1.63 | 2.97 | 1.63 |
| | ATSE (*c=1*) | 0.49 | 1.32 | 0.64 | 2.76 | 2.75 | 2.85 | 2.81 | 3.05 | 2.92 |
| | ATSE (*c=4*) | 0.51 | 1.31 | 0.67 | 2.83 | 2.83 | 2.90 | 2.88 | 3.12 | 2.98 |
| | ATSE (*c=8*) | 0.52 | 1.25 | 0.71 | 3.05 | 3.05 | 3.04 | 3.09 | 3.30 | 3.12 |
| | ATSE (*c=4*) adj. for PFS | -0.10 | 0.29 | 0.01 | 3.38 | 3.38 | 3.36 | 3.38 | 3.39 | 3.36 |
| **Scenario 16** *High treatment effect & High switching proportion,* $N_{EC}$=1000 | Oracle | 0.00 | 0.00 | 0.00 | 3.51 | 3.51 | 3.51 | 3.51 | 3.51 | 3.51 |
| | ITT | 7.59 | 7.59 | 6.66 | 3.43 | 3.43 | 3.46 | 8.32 | 8.32 | 7.50 |
| | TSE | 1.22 | 1.22 | 1.63 | 5.35 | 5.35 | 4.78 | 5.48 | 5.48 | 5.04 |
| | ECA | 0.00 | 2.51 | 0.00 | 1.64 | 1.63 | 1.64 | 1.64 | 2.99 | 1.64 |
| | ATSE (*c=1*) | 0.97 | 1.83 | 1.03 | 2.82 | 2.75 | 2.97 | 2.98 | 3.30 | 3.14 |
| | ATSE (*c=4*) | 0.93 | 1.77 | 1.02 | 2.86 | 2.83 | 3.05 | 3.01 | 3.34 | 3.22 |
| | ATSE (*c=8*) | 1.00 | 1.77 | 1.11 | 3.07 | 3.04 | 3.19 | 3.23 | 3.52 | 3.38 |
| | ATSE (*c=4*) adj. for PFS | -0.46 | -0.05 | -0.34 | 3.55 | 3.53 | 3.57 | 3.58 | 3.53 | 3.58 |

**Table 5. Simulation study results for Scenarios 13-16 (moderate censoring; large external control, $N_{EC}$=1000).**

**R code for simulating a subject for the simulation study.**

```
####################################################################################
sim_function <- function(trt = 0,
                         switching = 1,
                         high_switching = 0,
                         enddate = 546,
                         pmix = 0.5,
                         lambda1 = 12.5,
                         lambda2 = 10,
                         gamma1 = 2,
                         gamma2 = 3,
                         delta1 = 2,
                         delta2 = 1,
                         delta3 = -0.3,
                         omega = 1.1,
                         pr_of_poorprog = 0.5,
                         bias = FALSE,
                         unmeasured_confounder = FALSE){

 # The probability of switching was set at 0.8 for patients in the poor prognosis group,
 # and at ----0.3---- (slightly different than the 0.2 used by Latimer)
 # for patients in the good prognosis group in scenarios with a moderate switching
 # proportion, and at 0.9 and 0.6 for poor and good prognosis patients respectively
 # in scenarios with a high switching proportion.

 # Probability of poor prognosis 0.5
 badprog <- sample(c(0,1), prob=c(1-pr_of_poorprog, pr_of_poorprog))[1]

 # Probability of confounder 0.5
 confounder <- sample(c(0,1), prob=c(0.5-bias, 0.5+bias))[1]

 if(!unmeasured_confounder) {
  if(switching==0){
   prob_of_switching <- 0
  }

  if(switching==1){
   if(!high_switching){
    prob_of_switching <- 0.8*badprog + 0.3*(1-badprog) }

   if(high_switching){
    prob_of_switching <- 0.9*badprog + 0.6*(1-badprog) }
```

```
  }
}


if(unmeasured_confounder) {
  if(switching==0){
    prob_of_switching <- 0
  }

  if(switching==1){
    if(!high_switching){
      prob_of_switching <- 0.8*badprog + 0.3*(1-badprog) - 0.2*confounder}

    if(high_switching){
      prob_of_switching <- 0.9*badprog + 0.6*(1-badprog) - 0.2*confounder}
  }
}


# Disease progression times were simulated to equal OS times multiplied
# by a random draw from a Beta(5,10) distribution, so that, on average,
# the duration of an individual's PFS is equal to about one third of their OS.
TTP_multiplier <- (rbeta(1,5,10))

f <- function(t) (1-(pmix*exp(-lambda1*t^(gamma1)) +
                     (1-pmix)*exp(-lambda2*t^(gamma2)))^exp(delta1*trt +
                        delta2*badprog + delta3*confounder))
f.inv <- inverse(f,lower=0,upper=1000000)

u_sample <- runif(1, 0, 1)

OS <- (c(unlist(lapply(u_sample, f.inv))))

PFS_exact  <- OS * TTP_multiplier
PFS_visit  <- min( c(OS,ceiling(PFS_exact/12)*12) )
PPS_visit <- OS-PFS_visit
PPS_visit_true <- PPS_visit

# switch?
switch_indicator <- 0
if(trt==0){
  switch_indicator <- sample(c(0,1), 1, prob= c(1-prob_of_switching, prob_of_switching))
}

PPS_visit_switch <- PPS_visit
if(switch_indicator){
  PPS_visit_switch <- PPS_visit*omega
}
```

```
  OSswitch <- PFS_visit + PPS_visit_switch

  OSswitchstatus <- as.numeric(OSswitch < enddate)
  OSswitch <- min(c(OSswitch, enddate))

  OSstatus <- as.numeric(OS < enddate)
  OS <- min(c(OS, enddate))

  TTPexactstatus <- as.numeric(PFS_exact  < enddate)
  TTPexact <- min(c(PFS_exact, enddate))

  TTPvisitstatus <- as.numeric(PFS_visit  < enddate)
  TTPvisit <- min(c(PFS_visit, enddate))

  ## PPS is the PPS with the impact of switching
  PPSstatus <- 1
  PPS <- OSswitch - TTPvisit

  if(TTPvisitstatus==0){
    PPSstatus <- 0
    PPS <- 0
  }

  if(OSswitchstatus==0){
    PPSstatus <- 0
    PPS <- OSswitch - TTPvisit
  }

  return(list(
    OS = unname(OS),
    OSstatus = unname(OSstatus),
    PPS = unname(PPS),
    PPSstatus = unname(PPSstatus),
    OSswitch = unname(OSswitch),
    OSswitchstatus = unname(OSswitchstatus),
    TTPexact = unname(TTPexact),
    TTPexactstatus = unname(TTPexactstatus),
    TTP = unname(TTPvisit),
    TTPstatus = unname(TTPvisitstatus),
    SWITCH= unname(switch_indicator),
    x1 = unname(badprog),
    censor_time = enddate))
}
s
```